\documentstyle[12pt,psfig]{article}
\def\reference{\parskip 0pt\par\noindent\hangindent 0.5 truecm}

\def\degr{\hbox{$^\circ$}}

\def\farcm{\hbox{$.\mkern-4mu^\prime$}}

\begin{document}
%
%
\title{1245+676 --- A CSO/GPS Source being an Extreme Case of a Double-Double
Structure}
%


\author{Andrzej Marecki$^{1}$,
 Peter D. Barthel$^{2}$,
 Antonis Polatidis$^{3}$ \and and
 Izabela Owsianik$^{3}$
} 

\date{}
\maketitle

{\center
$^1$ Toru\'n Centre for Astronomy, Nicholas Copernicus University, Toru\'n, Poland\\[3mm]
$^2$ Kapteyn Institute, Department of Astronomy, Groningen University, Netherlands\\[3mm]
$^3$ Max-Planck-Institut f\"ur Radioastronomie, Bonn, Germany\\[3mm]
}

%
\begin{abstract}

AGN with the so-called `double-double' radio structure have been interpreted as restarted AGN where the inner
structure is a manifestation of a new phase of activity which happened to begin before the outer radio lobes
resulting from the previous one had faded completely. The radio galaxy 1245+676 is an extreme example of such a
double-double object --- its outer structure, measuring $970\,h^{-1}$\,kpc, is five orders of magnitude larger
than the $9.6\,h^{-1}$\,pc inner one. We present a series of VLBI observations of the core of 1245+676 which
appears to be a compact symmetric object (CSO). We have detected the motion of the CSO's lobes, measured its
velocity, and inferred the kinematic age of that structure.

\end{abstract}


\bigskip

\section{Objects with the `Double-Double' Radio Structure}

Radio galaxies which are not beamed toward the observer are normally perceived as double structures. The majority
of them are large scale objects (LSOs) with angular sizes ranging from several arcseconds to several arcminutes.
These angular sizes translate to large linear sizes of the order of $10^5-10^6$\,pc; galaxies with sizes $>1$\,Mpc
are labelled `giant radio galaxies' (GRGs). A very interesting exception to this (relatively simple) picture
exists however: a few LSOs, which are also double at first sight, turn out to have so-called double-double
structure, radio galaxy 1450+333 (Schoenmakers~et~al.~2000) being a prime example.

LSO radio lobes are powered by central engines for a maximum of
approximately $10^7$~years (Alexander \& Leahy~1987; Liu, Pooley,
\& Riley~1992). If the nuclear energy supply stops, the extended
radio structure will stop growing, its luminosity drops and the
spectrum gradually gets steeper and steeper because of radiation
and expansion losses. Komissarov \& Gubanov (1994) estimate
fade-away time scales of several $10^7$ years which is comparable
to the timescale of the activity itself.

After that time the radio structure should in principle disappear, however there are a number of known mechanisms
of restarting activity. For example, Hatziminaoglou, Siemiginowska, \&~Elvis~(2001) elaborated a theory of
thermal--viscous instabilities in the accretion disks of supermassive black holes (SMBHs). It predicts that the
activity is recurrent and the length of the activity phase as well as the timescale of the activity re-occurrence
is controlled by the mass of the SMBH. In favourable circumstances a new phase of activity can start before the
radio lobes resulting from the previous one have faded completely. This would lead to double-double radio
structure.

\section{Inner and Outer Radio Structures of 1245+676}

\begin{figure}[ht]
\begin{center}
\psfig{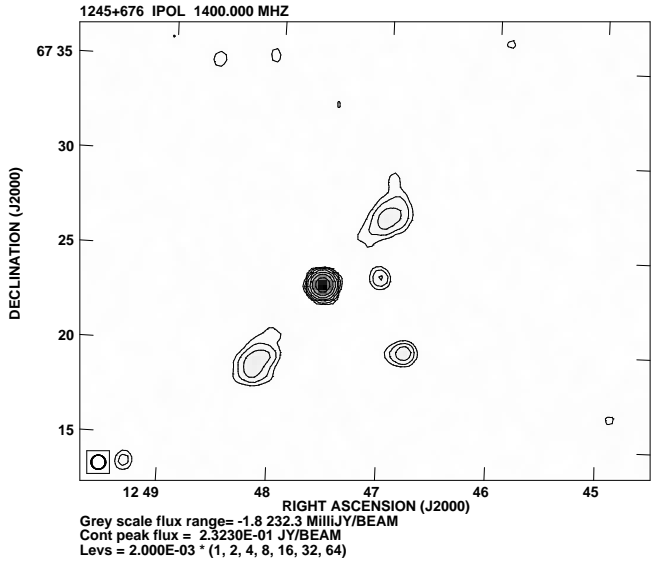}
\caption{NVSS image of the whole structure of the radio galaxy
1245+676. The radio galaxy stretches SE--NW over 12\farcm4 ---
also two compact field sources are seen.}
\label{NVSS}            
\end{center}
\end{figure}

1245+676 is a classical double-lobed radio galaxy redshifted to $z=0.1073$. Its angular size amounts to
$12\farcm4$ (Lara~et~al.~2001) which is equivalent to $970\,h^{-1}$\,kpc (assuming $q_0=0.5$). Figure \ref{NVSS}
shows the NVSS map of 1245+676. The central component is noticeably strong. Indeed, according to
Lara~et~al.~(2001) $2/3$ of the whole flux at 1.4\,GHz is emitted by the core. Moreover the core has a
gigahertz-peaked spectrum (GPS) --- see for example the spectrum shown in Marecki~et~al.~(1999).

\begin{figure}[t]
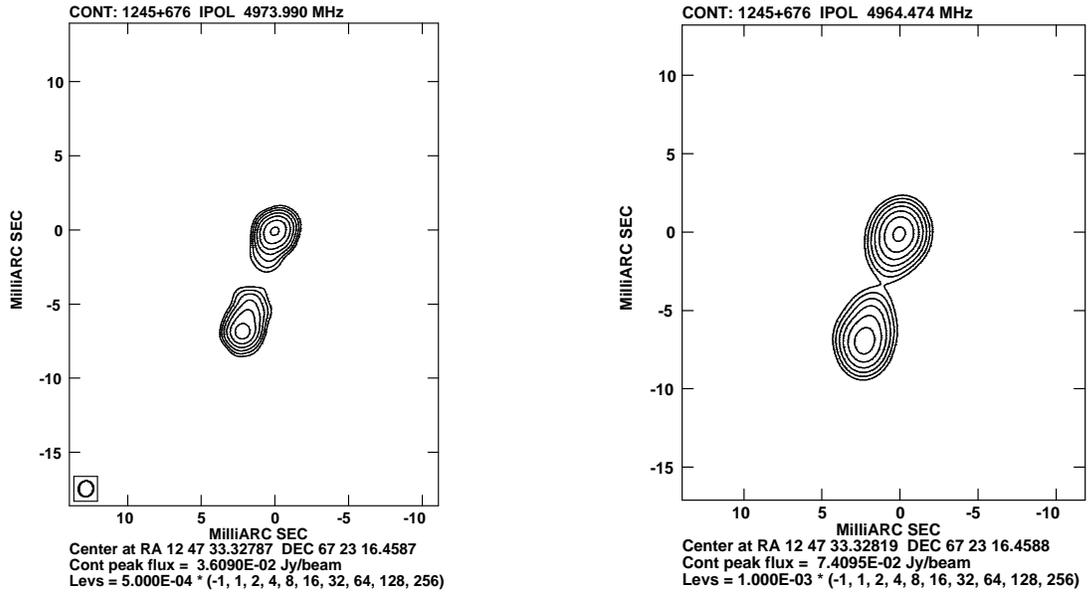

\begin{center}
\hbox{ \psfig{file=1245+676_PDB_6cm.ps,height=8.5cm}\hspace{2.0cm}
\psfig{file=1245+676_AM_6cm.ps,height=8.5cm} } \caption{VLBI
images of 1245+676 at 5\,GHz for epochs 1991 and 1998.}
\label{6cm}
\end{center}
\end{figure}

\begin{figure}
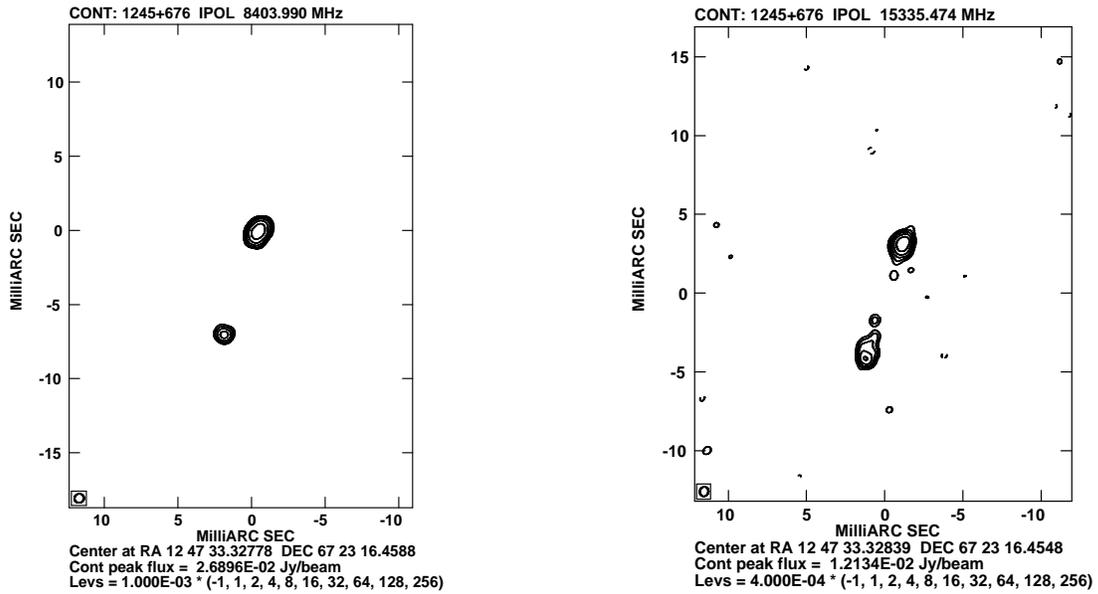

\begin{center}
\hbox{
\psfig{file=1245+676_PDB_3.6cm.ps,height=8.5cm}\hspace{2.5cm}
\psfig{file=1245+676_AM_2cm.ps,height=8.5cm} } \caption{VLBI
images of 1245+676 at 8.4\,GHz (epoch 1991) and 15\,GHz (epoch
1998).} \label{hifreq}
\end{center}
\end{figure}

A number of VLBI observations of the core of 1245+676 have been conducted in different epochs. More technical
details about these observations will be given elsewhere. In Figure \ref{6cm} we present a selection of resulting
5\,GHz images; in Figure \ref{hifreq} we show the 8.4\,GHz map made in 1991 and the most detailed VLBI map of the
1245+676 core observed in 1998 at 15\,GHz. These maps clearly show that the `core' is actually a compact symmetric
object (CSO) with a double-lobed inner structure. Its size is only $9.6\,h^{-1}$\,pc. This means 1245+676 is an
extreme case of a double-double object --- its outer structure is five orders of magnitude larger than the inner
one.

We are confident that the core of 1245+676 is indeed a CSO because
based on our dual frequency (5 and 15\,GHz) quasi-simultaneous
VLBA observations made on 6~September~1998 we were able to
calculate the spectral indices of the milliarcsecond scale
components. They amount to $\alpha_{North}=-1.03$ and
$\alpha_{South}=-1.23$ so they are both steep and roughly equal.
We can therefore safely rule out a possibility they are not lobes
but, for example, a core and a jet. (This approach is similar to
that adopted by Peck \&~Taylor~2000.)

We note a modest misalignment between the parsec scale ($PA=-19\degr$) and
the megaparsec scale ($PA=-50\degr$) which may be attributed to precession
of SMBH and the accretion disk.

\section{Expansion Velocity and the Kinematic Age}

Given the substantial time baseline between the 5\,GHz
observations we attempted the measurement of the lobes' expansion
velocity. We used different software tools from AIPS and Difmap
packages, and the final expansion figure is an average of the
measurements obtained. It amounts to $0.0349\pm0.0018\,{\mathrm
mas}\,{\mathrm yr^{-1}}$, which is equivalent to
$v_{exp}=0.164\pm0.008\,h^{-1}c$. The inferred kinematic age of
the CSO structure is 191~years.

\section*{References}


\reference Alexander, P., \& Leahy, J.P.\ 1987, MNRAS, 225, 1
\reference Hatziminaoglou, E., Siemiginowska, A., \& Elvis, M.\
2001, ApJ, 547, 90

\reference Komissarov, S.S., \& Gubanov, A.G.\ 1994, A\&A, 285, 27

\reference Lara, L., Cotton, W.D., Feretti, L., Giovannini, G.,
Marcaide, J.M., M\'arquez, I., \& Venturi, T.\ 2001, A\&A, 370,
409

\reference Liu, R., Pooley, G.G., \& Riley, J.M.\ 1992, MNRAS,
257, 545

\reference Marecki, A., Falcke, H., Niezgoda, J., Garrington,
S.T., \& Patnaik, A.R.\ 1999, A\&AS, 135, 273

\reference Peck, A.B., \& Taylor, G.B.\ 2000, ApJ, 534, 90

\reference Schoenmakers, A.P., de Bruyn, A.G., R\"ottgering,
H.J.A., van der Laan, H., \& Kaiser,~C.R.\ 2000, MNRAS, 315, 371

\vspace{20cm}

\end{document}